\documentclass{article}

\usepackage[centertags]{amsmath}
\usepackage{chicago,amssymb,amsfonts,eucal,oldgerm,fancyhdr}
\usepackage[polutonikogreek,french,german,english]{babel}

\newcommand{\skipline}[1][1]{\vspace*{#1\baselineskip}}

\newcommand{\english}{\selectlanguage{english}}

\newtheorem{theorem}{Theorem}[subsection]

\newtheorem{condition}[theorem]{Condition}

\newtheorem{defn}[theorem]{Definition}
\newtheorem{lemma}[theorem]{Lemma}

\newtheorem{prop}[theorem]{Proposition}

\newcommand{\resetcounters}{%
  \setcounter{equation}{0}%
  \setcounter{theorem}{0}%
  \setcounter{figure}{0}%
}

\newcommand{\resetsec}{%
  \resetcounters%
  \renewcommand{\thetheorem}{\arabic{section}.\arabic{theorem}}%
  \renewcommand{\theequation}{\arabic{section}.\arabic{equation}}%
}

\renewcommand{\thetheorem}{\arabic{section}.\arabic{theorem}}%
\renewcommand{\theequation}{\arabic{section}.\arabic{equation}}%

\newcommand{\half}{\frac{1}{2}}

\title{On Tensorial Concomitants and the Non-Existence of a
  Gravitational Stress-Energy Tensor\footnote{This paper was submitted
    (February 2012) to \emph{General Relativity and Gravitation}.}}

\author{Erik Curiel\footnote{I thank Robert Geroch for many
    stimulating conversations in which the seeds of several of the
    paper's ideas were germinated and, in some cases, fully cultivated
    to fruition.  I also thank David Malament for helpful
    conversations on the principle of equivalence and on gravitational
    energy.  Finally, I am grateful to Ted Jacobson for commenting on
    an earlier draft.  Author's email: erik@strangebeautiful.com;
    address: Rotman Institute of Philosophy, University of Western
    Ontario, London, Ontario N6A 5B8, Canada; +1-519-661-2111}}

\date{February, 2012}

\begin{document}
\english

\maketitle

\begin{quote}
  \begin{center}
    \textbf{ABSTRACT}
  \end{center}
  The question of the existence of gravitational stress-energy in
  general relativity has exercised investigators in the field since
  the inception of the theory.  Folklore has it that no adequate
  definition of a localized gravitational stress-energetic quantity
  can be given.  Most arguments to that effect invoke one version or
  another of the Principle of Equivalence.  I argue that not only are
  such arguments of necessity vague and hand-waving but, worse, are
  beside the point and do not address the heart of the issue.  Based
  on a novel analysis of what it may mean for one tensor to depend in
  the proper way on another, I prove that, under certain natural
  conditions, there can be no tensor whose interpretation could be
  that it represents gravitational stress-energy in general
  relativity.  It follows that gravitational energy, such as it is in
  general relativity, is necessarily non-local.  Along the way, I
  prove a result of some interest in own right about the structure of
  the associated jet bundles of the bundle of Lorentz metrics over
  spacetime.
\end{quote}

\skipline

\noindent \textbf{Keywords:} gravitational energy; stress-energy
tensors; concomitants; jet bundles; principle of equivalence

\tableofcontents

\section{Gravitational Energy in General Relativity}
\label{sec:intro}

There seems to be in general relativity no satisfactory, localized
representation of a quantity whose natural interpretation would be
``gravitational (stress-)energy''.  The only physically unquestionable
expressions of energetic quantities associated solely with the
``gravitational field'' we know of in general relativity are
quantities derived by integration over non-trivial volumes in
spacetimes satisfying any of a number of special
conditions.\footnote{\citeN[pp.~271--272]{weyl-space-time-matter} was
  perhaps the first to grasp this point with real clarity.  See also
  \citeN{dirac-grav-energy}.
  \citeN[pp.~104--105]{schrodinger-st-struc} gives a particularly
  clear, concise statement of the relation between the fact that the
  known energetic, gravitational quantities are non-tensorial and the
  fact that integration over them can be expected to yield integral
  conservation laws only under restricted conditions.}  These
quantities, moreover, tend to be non-tensorial in character.  In other
words, these are strictly non-local quantities, in the precise sense
that they are not represented by invariant geometrical objects defined
at individual spacetime points (such as tensors or scalars).

This puzzle about the character and status of gravitational energy
emerged simultaneously with the discovery of the theory
itself.\footnote{The first pseudo-tensorial entity proposed to
  represent gravitational stress-energy dates back to
  \citeN{einstein-gr}, the paper in which he first proposed the final
  form of the theory.}  The problems raised by the seeming
non-localizability of gravitational energy had a profound, immediate
effect on subsequent research.  For example, it was directly
responsible for Hilbert's request to Noether that she investigate
conservation laws in a quite general setting, the work that led to her
famous results relating symmetries and conservation
laws.\footnote{See, \emph{e}.\emph{g}.,
  \citeN{brading-energy-cons-gr}.}

Almost all discussions of gravitational energy in general relativity,
however, dating back to those earliest debates, have been plagued by
vagueness and lack of precision.  The main result of this paper
addresses the issue head-on in a precise and rigorous way.  Based on
an analysis of what it may mean for one tensor to depend in the proper
way on another, I prove that, under certain natural conditions, there
can be no tensor whose interpretation could be that it represents
gravitational stress-energy in general relativity.  It follows that
gravitational energy, such as it is in general relativity, is
necessarily non-local.  Along the way, I prove a result of some
interest in own right about the structure of the associated first two
jet bundles of the bundle of Lorentz metrics over spacetime.

\section{The Principle of Equivalence:  A Bad Argument}
\label{sec:princ_equiv}
\resetsec

The most popular heuristic argument used to attempt to show that
gravitational energy either does not exist at all or does exist but
cannot be localized invokes the ``Principle of Equivalence''.
\citeN[p.~399]{choquet83}, for example, puts the argument like this:
\begin{quote}
  This `non local' character of gravitational energy is in fact
  obvious from a formulation of the equivalence principle which says
  that the gravitational field appears as non existent to one observer
  in free fall.  It is, mathematically, a consequence of the fact that
  the pseudo-riemannian connexion which represents the gravitational
  field can always be made to vanish along a given curve by a change
  of coordinates.
\end{quote}
\citeN[pp.~469-70]{goldberg80} makes almost exactly the same argument,
though he draws the conclusion in a slightly more explicit
fashion:\footnote{Goldberg's formulation of the argument exhibits a
  feature common in the many instances of it I have found in the
  literature, the conclusion that a local gravitational energy
  \emph{scalar} density does not exist and not that a gravitational
  stress-energy tensor does not exist.  Perhaps one could imagine
  having a well-defined scalar energy density of a field in the
  absence of a well-defined stress-energy tensor for that field,
  though I cannot myself see any way to represent such an idea in
  general relativity.  (Note that if one could, this would appear to
  be a violation of the thermodynamic principle that all energy is
  equivalent in character, in the sense that any one form can always
  in principle be tranformed into any other form, since all other
  forms have a stress-energy tensor as their fundamental
  representation.)}
\begin{quote} [I]n Minkowski space any meaningful energy density
  should be zero.  But a general space-time can be made to appear
  Minkowskian along an arbitrary geodesic.  As a result, any
  nontensorial `energy density' can be made to be zero along an
  arbitrary geodesic and, therefore, has no invariant meaning.
\end{quote}
\citeN[pp.~135-6]{trautman76} has also made essentially the same
argument.  In fact, the making of this argument seems to be something
of a shared mannerism among physicists who discuss energy in general
relativity; it is difficult to find an article on the topic in which
it is not at least alluded to.\footnote{\citeN{bondi62},
  \citeN{penrose66} and \citeN{geroch73} are notable exceptions.  I
  take their discussions as models of how one should discuss energetic
  phenomena in the presence of gravitational fields.}

The argument has a fundamental flaw.  It assumes that, if there is
such a thing as localized gravitational energy or stress-energy, it
can depend only on ``first derivatives of the metric''---that those
first derivatives encode all information about the ``gravitational
field'' relevant to stress-energy.  But that seems wrong on the face
of it.  If there is such a thing as a localized gravitational
energetic quantity, then surely it depends on the curvature of
spacetime and not on the affine connection (or, more precisely, it
depends on the affine connection at least in so far as it depends on
the curvature), for any energy one can envision transferring from the
gravitational field to another type of system in a different form
(\emph{e}.\emph{g}., as heat or a spray of fundamental particles) in
general relativity must at bottom be based on geodesic
deviation,\footnote{\citeN{penrose66} and \citeN{asktekar-penrose90}
  implicitly rely on the same idea to very fruitful effect.} and so
must be determined by the value of the Riemann tensor at a point, not
by the value of the affine connection at a point or even along a
curve.  There is no solution to the Einstein field-equation that
corresponds in any natural way to the intuitive Newtonian idea of a
constant gravitational field, \emph{i}.\emph{e}., one without geodesic
deviation; that, however, would be the only sort of field that one
could envision even being tempted to ascribe gravitational energy to
in the absence of geodesic deviation, and that attribution is
problematic even in Newtonian theory.  Indeed, a spacetime has no
geodesic deviation if and only if it is (locally) isometric to
Minkowski spacetime, which we surely want to say is the unique
spacetime to have vanishing gravitational energy, if one can make such
a statement precise in the first place.

An obvious criticism of my response to the standard line, related to a
popular refinement of the argument given for the non-existence or
non-locality of gravitational energetic quantities, is that it would
make gravitational stress-energy depend on second-order partial
derivatives of the field potential (the metric, so comprehended by
analogy with the potential in Newtonian theory), whereas all other
known forms of stress-energy depend only on terms quadratic in the
first partial derivatives of the field potential.  To be more precise,
the argument runs like this:
\begin{quote}
  One can make precise the sense in which Newtonian gravitational
  theory is the `weak-field' limit of general
  relativity.\footnote{See, \emph{e.g}., \citeN{malament86}.}  In this
  limit, it is clear that the metric field plays roughly the role in
  general relativity that the scalar potential $\phi$ does in
  Newtonian theory.  In Newtonian theory, bracketing certain technical
  questions about boundary conditions, there is a more or less
  well-defined energy density of the gravitational field, proportional
  to $( \nabla \phi )^2$.  One might expect, therefore, based on some
  sort of continuity argument, or just on the strength of the analogy
  itself, that any local representation of gravitational energy in
  general relativity ought to be a ``quadratic function of the first
  partials of the metric''.\footnote{In this light, it is interesting
    to note that gravitational energy pseudo-tensors do tend to be
    quadratic in the first-order partials of the metric.}  The
  stress-energy tensor of no other field, moreover, is higher than
  first-order in the partials of the field potential, so surely
  gravity cannot be different.  No invariant quantity at a point can
  be constructed using only the first partials of the metric, however,
  so there can be no scalar or tensorial representation of
  gravitational energy in general relativity.
\end{quote}
(No writer I know makes the argument exactly in this form; it is just
the clearest, most concise version I can come up with myself.)  As
\citeN[p.\ 178]{pauli21} forcefully argued, however, there can be no
\emph{physical} argument against the possibility that gravitational
energy depends on second derivatives of the metric; the argument above
certainly provides none.  Just because the energy of all other known
fields have the same form in no way implies that a localized
gravitational energy in general relativity, if there is such a thing,
ought to have that form as well.  Gravity is too different a field
from others for such a bare assertion to carry any weight.  As I
explain at the end of \S\ref{sec:conds}, moreover, a proper
understanding of tensorial concomitants reveals that an expression
linear in second partial derivatives is in the event equivalent in the
relevant sense to one quadratic in first order partials.  This
illustrates how misleading the analogy with Newtonian gravity can be.

\section{Geometric Fiber Bundles and Concomitants}
\label{sec:concomitants}
\resetsec

\begin{quote}
  The introduction of a coordinate system to geometry is an act of
  violence.
  \skipline[-.5]
  \begin{flushright}
    Hermann Weyl \\
    \emph{Philosophy of Mathematics and Natural Science}
  \end{flushright}
\end{quote}

I have argued that, if there is an object that deserves to be thought
of as the representation of gravitational stress-energy of in general
relativity, then it ought to depend on the Riemann curvature tensor.
Since there is no obvious mathematical sense in which a general
mathematical structure can ``depend'' on a tensor, the first task is
to say what exactly this could mean.  I will call a mathematical
structure on a manifold that depends in the appropriate fashion on
another structure on the manifold, or set of others, a
\emph{concomitant} of it (or them).

The reason I am inquiring into the possibility of a concomitant in the
first place, when the question is the possible existence of a
representation of gravitational stress-energy tensor, is a simple one.
What is wanted is an expression for gravitational energy that does not
depend for its formulation on the particulars of the spacetime, just
as the expression for the kinetic energy of a particle in classical
physics does not depend on the particular interactions one imagines
the particle to be experiencing with its environment, and just as the
stress-energy tensor for a Maxwell field can be calculated in any
spacetime in which there is a Maxwell field, irrespective of the
particulars of the spacetime, in contradistinction to the definitions
of all known expressions for gravitational energy in general
relativity (\emph{e}.\emph{g}., the ADM mass, which can be defined
only in asymptotically flat spacetimes).  If there is a well-formed
expression for gravitational stress-energy, then one should be able in
principle to calculate it whenever there are gravitational phenomena,
which is to say, in any spacetime whatsoever---it should be a
\emph{function} of some set of geometrical objects associated with the
curvature in that spacetime, in some appropriately generalized sense
of `function'.  This idea is what a concomitant is supposed to
capture.

As near as I can make out, the term `concomitant' and the general idea
of the thing is due to
Schouten.\footnote{\label{fn:schouten-lovelock}See
  \citeN[p.~15]{schouten23/54}, though of course he used the German
  \emph{Komitant}.  The specific idea of proving the uniqueness of a
  tensor that ``depends'' on another tensor, and satisfies a few
  collateral conditions, dates back at least to
  \citeN[pp.~315-18]{weyl-space-time-matter} and \citeN{cartan22}.  In
  fact, Weyl proved that, in any spacetime, the only two-index
  symmetric covariant tensors one can construct at a point, using only
  algebraic combinations of the components of the metric and its first
  two partial derivatives in a coordinate system at that point, that
  are at most linear in the second derivatives of the metric, are
  linear combinations of the Ricci curvature tensor, the scalar
  curvature times the metric and the metric itself.  In particular,
  the only such divergence-free tensors one can construct at a point
  are linear combinations of the Einstein tensor and the metric with
  constant coefficients.  Using Schouten's definition of a
  concomitant, \citeN{lovelock72} proved the following theorem:
  \begin{quote}
    \emph{Let $(\mathcal{M},\; g_{ab})$ be a spacetime.  In a
      coordinate neighborhood of a point $p\in \mathcal{M}$, let
      $\Theta_{\alpha \beta}$ be the components of a tensor
      concomitant of $\{g_{\lambda \mu} ; \; g_{\lambda \mu,\nu} ; \;
      g_{\lambda \mu , \nu \rho} \}$ such that}
    \[
    \nabla^n \Theta_{nb} = 0.
    \]
    \emph{Then}
    \[
    \Theta_{ab} = r G_{ab} + q g_{ab},
    \]
    \emph{where $G_{ab}$ is the Einstein tensor and $q$ and $r$ are
      constants.}
  \end{quote}
  This is a much stronger result in several ways than Weyl and Cartan
  had been able to attain: one has a more generalized notion of
  concomitant than algebraic combination of coordinate components; one
  does not demand that $\Theta_{ab}$ be symmetric; and most
  strikingly, one does not demand that $\Theta_{ab}$ be at most linear
  in the second-order partial derivatives of the metric components.}
The definition Schouten proposed---the only one I know of in the
literature---is expressed in terms of coordinates: depending on what
sort of concomitant one was dealing with, the components of the object
had to satisfy various conditions of covariance under certain classes
of coordinate transformations.  This makes it not only unwieldy in
practice and inelegant, but, more important, it makes it difficult to
discern what of intrinsic physical significance is encoded in the
relation of being a concomitant in particular cases.  Schouten's
covariance conditions translate into a set of partial differential
equations in a particular coordinate system, which even in relatively
straightforward cases turn out to be forbiddingly
complicated.\footnote{For a good example of just how hairy these
  conditions can be, see \citeN[p.~350]{duplessis69} for a complete
  set written out explicitly in the case of two covariant-index
  tensorial second-order differential concomitants of a Lorentz
  metric.}  It is almost impossible to determine anything of the
general properties of the set of a particular kind of concomitant of a
particular object by looking at these equations.  I suspect that it is
because these conditions are so complex, difficult and opaque that use
is very rarely made of concomitants in arguments about spacetime
structure in general relativity.  This is a shame, for the idea is, I
think, potentially rich, and so calls out for an invariant
formulation.

I use the machinery of fiber bundles to characterize the idea of a
concomitant in invariant terms.  I give a (brief) explicit formulation
of the machinery, because the one I rely on is non-standard.  (We
assume from hereon that all relevant structures, mappings,
\emph{etc}., are smooth.  Nothing is lost by the assumption and it
will simplify exposition.  All constructions and proofs can easily be
generalized to the case of topological spaces and continuous
structures.)
\begin{defn}
  \label{defn:fiberbundle}
  A \emph{fiber} \emph{bundle} $\mathfrak{B}$ is an ordered triplet,
  $\mathfrak{B} \equiv (\mathcal{B} , \space \mathcal{M} , \space
  \pi)$, such that:
  \begin{enumerate}
    \addtolength{\itemindent}{2em} 
      \item[\bf{FB1}.] $\mathcal{B}$ is a differential manifold
      \item[\bf{FB2}.] $\mathcal{M}$ is a differential manifold
      \item[\bf{FB3}.] $\pi : \mathcal{B} \rightarrow \mathcal{M}$ is
    smooth and onto
      \item[\bf{FB4}.] For every $q,p \in \mathcal{M}$, $\pi^{-1} ( q
    )$ is diffeomorphic to $\pi^{-1} ( p )$ (as submanifolds of
    $\mathcal{B}$)
      \item[\bf{FB5}.] $\mathcal{B}$ has a locally trivial product
    structure, in the sense that for each $q \in \mathcal{M}$ there is
    a neighborhood $U \ni q$ and a diffeomorphism $\zeta : \pi^{-1}
    [U] \rightarrow U \times \pi^{-1} (q)$ such that the action of
    $\pi$ commutes with the action of $\zeta$ followed by projection
    on the first factor.
  \end{enumerate}
\end{defn}
$\mathcal{B}$ is the \emph{bundle space}, $\mathcal{M}$ the \emph{base
  space}, $\pi$ the \emph{projection} and $\pi^{-1} ( q )$ the
\emph{fiber} over $q$.  By a convenient, conventional abuse of
terminology, I will sometimes call $\mathcal{B}$ itself `the fiber
bundle' (or `the bundle' for short).  A \emph{cross-section} $\kappa$
is a smooth map from $\mathcal{M}$ into $\mathcal{B}$ such that $\pi
(\kappa (q)) = q$, for all $q$ in the mapping's domain.

This definition of a fiber bundle is non-standard in so far as no
group action on the fibers is fixed from the start; this implies that
no correlation between diffeomorphisms of the base space and
diffeomorphisms of the bundle space is fixed.\footnote{See,
  \emph{e}.\emph{g}., \citeN{steenrod51} for the traditional
  definition and the way that a fixed group action on the fibers
  induces a correlation between diffeomorphisms on the bundle space
  and those on the base space.}  One must fix that explicitly.  On the
view I advocate, the geometric character of the objects represented by
the bundle arises arises not from the group action directly, but only
after the explicit fixation of a correlation between diffeomorphisms
on the base space with those on the bundle space---only after, that
is, one fixes how a diffeomorphism on the base space induces one on
the bundle.  For example, depending on how one decides that a
diffeomorphism on the base space ought to induce a diffeomorphism on
the bundle over it whose fibers consist of 1-dimensional vector
spaces, one will ascribe to the objects of the bundle the character
either of ordinary scalars or of $n$-forms (where $n$ is the dimension
of the base space).  The idea is that the diffeomorphisms induced on
the bundle space then implicitly define the group action on the fibers
appropriate for the required sort of object.\footnote{I will not work
  out the details of how this comes about here, as they are not needed
  for the arguments of the paper.}

I call an appropriate mapping of diffeomorphisms on the base space to
those on the bundle space an \emph{induction}.  (I give a precise
definition in a moment.)  In this scheme, therefore, the induction
comes first conceptually, and the relation between diffeomorphisms on
the base space and those they induce on the bundle serves to fix the
fibers as spaces of \emph{geometric objects}, \emph{viz}., those whose
transformative properties are tied directly and intimately to those of
the ambient base space.  This way of thinking of fiber bundles is
perhaps not well suited to the traditional task of classifying
bundles, but it turns out to be just the thing on which to base a
perspicuous and useful definition of concomitant.  Although a
diffeomorphism on a base space will naturally induce a unique one on
certain types of fiber bundles over it, such as tensor bundles, in
general it will not.  There is not known, for instance, any natural
way to single out a map of diffeomorphisms of the base space into
those of a bundle over it whose fibers consist only of spinorial
objects.\footnote{See, \emph{e}.\emph{g}.,
  \citeN{penrose-rindler-spinors-st-1}.}  Inductions neatly handle
such problematic cases.

I turn now to making this intuitive discussion more precise.  A
diffeomorphism $\phi^\sharp$ of a bundle space $\mathcal{B}$ is
\emph{consistent} with $\phi$, a diffeomorphism of the base space
$\mathcal{M}$, if, for all $u \in \mathcal{B}$,
\[
\pi (\phi^\sharp (u)) = \phi (\pi (u))
\] 
For a general bundle, there will be scads of diffeomorphisms
consistent with a given diffeomorphism on the base space.  A way is
needed to fix a unique $\phi^\sharp$ consistent with a $\phi$ so that
a few obvious conditions are met.  For example, the identity
diffeomorphism on $\mathcal{M}$ ought to pick out the identity
diffeomorphism on $\mathcal{B}$.  More generally, if $\phi$ is a
diffeomorphism on $\mathcal{M}$ that is the identity on an open set $O
\subset \mathcal{M}$ and differs from the identity outside $O$, it
ought to be the case that the mapping picks out a $\phi^\sharp$ that
is the identity on $\pi^{-1} [O]$.  If this holds, we say that that
$\phi^\sharp$ is \emph{strongly consistent} with $\phi$.

Let $\mathfrak{D}_\mathcal{M}$ and $\mathfrak{D}_\mathcal{B}$ be,
respectively, the groups of diffeomorphisms on $\mathcal{M}$ and
$\mathcal{B}$ to themselves, respectively.  Define the set
\begin{center}
  $\mathfrak{D}^\sharp_\mathcal{B} = \{ \phi^\sharp \in
  \mathfrak{D}_\mathcal{B} : \: \exists \phi \in
  \mathfrak{D}_\mathcal{M}$ such that $\phi^\sharp$ is strongly
  consistent with $\phi \}$
\end{center}
It is simple to show that $\mathfrak{D}^\sharp_\mathcal{B}$ forms a
subgroup of $\mathfrak{D}_\mathcal{B}$.  This suggests 
\begin{defn}
  \label{defn:induction}
  An \emph{induction} is an injective homomorphism $\iota :
  \mathfrak{D}_\mathcal{M} \rightarrow
  \mathfrak{D}^\sharp_\mathcal{B}$.
\end{defn}
$\phi$ will be said to \emph{induce} $\phi^\sharp$ (under $\iota$) if
$\iota ( \phi ) = \phi^\sharp$.\footnote{In a more thorough treatment,
  one would characterize the way that the induction fixes a group
  action on the fibers, but we do not need to go into that for our
  purposes.}
\begin{defn}
  \label{defn:geom-bundle}
  A \emph{geometric fiber bundle} is an ordered quadruplet
  $(\mathcal{B}, \, \mathcal{M}, \, \pi, \, \iota)$ where
  \begin{enumerate}
    \addtolength{\itemindent}{3em} 
      \item[{\bf GFB1}.] $(\mathcal{B}, \, \mathcal{M}, \, \pi)$
    satisfies FB1-FB5
      \item[{\bf GFB2}.] $\iota$ is an induction
  \end{enumerate}
\end{defn} 
Geometric fiber bundles are the appropriate spaces to serve as the
domains and ranges of concomitant mappings.

Most of the fiber bundles one works with in physics are geometric
fiber bundles.  A tensor bundle $\mathcal{B}$, for example, is a fiber
bundle over a manifold $\mathcal{M}$ each of whose fibers is
diffeomorphic to the vector space of tensors of a particular index
structure over any point of the manifold; a basis for an atlas is
provided by the charts on $\mathcal{B}$ naturally induced from those
on $\mathcal{M}$ by the representation of tensors on $\mathcal{M}$ as
collections of components in $\mathcal{M}$'s coordinate systems.
There is a natural induction in this case, $\iota :
\mathfrak{D}_\mathcal{M} \rightarrow
\mathfrak{D}^\sharp_\mathfrak{B}$, fixed by the pull-back action of a
diffeomorphism $\phi$ of tensors on $\mathcal{M}$.  It is
straightforward to show that $\iota$ so defined is in fact an
induction.  Spinor bundles provide interesting examples of physically
important bundles that have no natural, unique inductions, though
there are classes of them.

We are finally in a position to define concomitants.  Let
$(\mathcal{B}_1, \, \mathcal{M}, \, \pi_1, \, \iota_1)$ and
$(\mathcal{B}_2, \, \mathcal{M}, \, \pi_2, \, \iota_2)$ be two
geometrical bundles with the same base space.\footnote{One can
  generalize the definition of concomitants to cover the case of
  bundles over different base spaces, but we do not need this here.}
\begin{defn}
  \label{def:concom}
  A mapping $\chi : \mathcal{B}_1 \rightarrow \mathcal{B}_2$ is a
  \emph{concomitant} if
  \[
  \chi (\iota_1 [\phi] (u_1)) = \iota_2 [\phi] (\chi (u_1))
  \]
  for all $u_1 \in \mathcal{B}_1$ and all $\phi \in
  \mathfrak{D}_\mathcal{M}$.
\end{defn}
In intuitive terms, a concomitant is a mapping between bundles that
commutes with the action of the induced diffeomorphisms that lend the
objects of the bundles their respective geometric characters.  It is
easy to see that $\chi$ must be fiber-preserving, in the sense that it
maps fibers of $\mathcal{B}_1$ to fibers of $\mathcal{B}_2$.  This
captures the idea that the dependence of the one type of object on the
other is strictly local; the respecting of the actions of
diffeomorphisms captures the idea that the mapping encodes an
invariant relation.

\section{Jet Bundles and Higher-Order Concomitants}
\label{sec:jet-bundles}
\resetsec

Just as with ordinary functions from one Euclidean space to another,
it seems plausible that the dependence encoded in a concomitant from
one geometric bundle to another may take into account not only the
value of the first geometrical structure at a point of the base space,
but also ``how that value is changing'' in a neighborhood of that
point, something like a generalized derivative of a geometrical
structure on a manifold.  The following construction is meant to
capture in a precise sense the idea of a generalized derivative in
such a way so as to make it easy to generalize the idea of a
concomitant to account for it.

Fix a geometric fibre bundle $(\mathcal{B}, \, \mathcal{M}, \, \pi, \,
\iota)$, and the space of its sections $\Gamma [\mathcal{B}]$.  Two
sections $\gamma, \eta : \mathcal{M} \rightarrow \mathcal{B}$
\emph{osculate to first-order} at $p \in \mathcal{M}$ if $T\gamma$ and
$T\eta$ (the differentials of the mappings) agree in their action on
$T_p \mathcal{M}$.  (They osculate to zeroth-order at $p$ if they map
$p$ to the same point in the domain.)  If $(x^i, \, v^\alpha)$ are
coordinates at the point $\gamma(q)$ adapted to the bundle structure
(as defined by the induction), then a coordinate representation of
this relation is:
\[
\left. \frac{\partial (v^\alpha \circ \gamma)}{\partial x^i} \right|_q
= \left. \frac{\partial (v^\alpha \circ \eta)}{\partial x^i} \right|_q
\]
for all $i \leq \dim(\mathcal{M})$ and $\alpha \leq
\dim(\pi^{-1}[q])$.  This defines an equivalence relation on $\Gamma
[\mathcal{B}]$.  A \emph{1-jet} with source $q$ and target
$\gamma(q)$, written `$j^1_q [\gamma]$', is such an equivalence class.
The set of all 1-jets,
\[
J^1 \mathcal{B} \equiv \bigcup_{q \in \mathcal{M}, \gamma \in \Gamma
  [\mathcal{B}]} j^1_q [\gamma]
\]
naturally inherits the structure of a differentiable manifold.  Let
$(\phi, \, U)$ be an adapted coordinate chart of $\mathcal{B}$ around
$\gamma(q)$, with the coordinate functions $(x^i, \, v^\alpha)$.  Then
the induced coordinate chart on $J^1 \mathcal{B}$ is $(\phi^1, \,
U^1)$ where
\begin{equation}
  \begin{split}
    U^1 \equiv \{ j^1_q [\gamma] \; | \; \gamma(q) \in U\}
  \end{split}
\end{equation}
and the coordinate functions associated with $\phi^1$ are $(x^i, \,
v^\alpha, \, v^\alpha_i)$, where
\begin{equation}
  \label{eq:coord-1jet}
  \begin{split}
    x^i(j^1_q [\gamma]) &\equiv x^i(q) \\
    v^\alpha(j^1_q [\gamma]) &\equiv v^\alpha (\gamma(q)) \\
    v^\alpha_i (j^1_q [\gamma]) &\equiv \left.  \frac{\partial
        (v^\alpha \circ \gamma)}{\partial x^i} \right|_q
  \end{split}
\end{equation}
where $\gamma$ is any member of $j^1_q [\gamma]$; this is well defined
since all members of $j^1_q [\gamma]$ agree on $\gamma(q)$ and
$\displaystyle \left. \frac{\partial (v^\alpha \circ \gamma)}{\partial
    x^i} \right|_q$ by definition.

One can naturally fibre $J^1 \mathcal{B}$ over $\mathcal{M}$.  The
\emph{source projection} $\sigma^1 : J^1 \mathcal{B} \rightarrow
\mathcal{M}$, defined by
\[
\sigma^1 (j^1_q [\gamma]) = q
\]
gives $J^1 \mathcal{B}$ the structure of a bundle space over the base
space $\mathcal{M}$, and in this case we write the bundle $(J^1
\mathcal{B}, \, \mathcal{M}, \, \sigma^1)$.  A section $\gamma$ of
$\mathcal{B}$ naturally gives rise to a section $j^1 [\gamma]$ of $J^1
\mathcal{B}$, the \emph{first-order prolongation} of that section:
\[
j^1 [\gamma] : \mathcal{M} \rightarrow \bigcup_{q \in \mathcal{M}}
j^1_q [\gamma]
\]
such that $\sigma_1 (j^1 [\gamma] (q)) = q$.  (We assume for the sake
of simplicity that global cross-sections exist; the modifications
required to treat local cross-sections are trivial.)

The points of $J^1 \mathcal{B}$ may be thought of as coordinate-free
representations of first-order Taylor expansions of sections of
$\mathcal{B}$.  To see this, consider the example of the trivial
bundle $(\mathcal{B}, \, \mathbb{R}^2, \, \pi)$ where $\mathcal{B}
\equiv \mathbb{R}^2 \times \mathbb{R}$ and $\pi$ is projection onto
the first factor.  Fix global coordinates $(x^1, \, x^2, \, v^1)$ on
$\mathcal{B}$, so that the induced (global) coordinates on $J^1
\mathcal{B}$ are $(x^1, \, x^2, \, v^1, \, v^1_1, \, v^1_2)$.
Then for any 1-jet $j^1_q [\gamma]$, define the inhomogenous linear
function $\hat{\gamma} : \mathbb{R}^2 \rightarrow \mathbb{R}$ by
\[
\hat{\gamma}(p) = v^1 (\gamma(p)) + v^1_1 (j^1_q [\gamma])(p_1 - q_1)
+ v^1_2 (j^1_q [\gamma])(p_2 - q_2)
\]
where $\gamma \in j^1_q [\gamma]$, and $p, q \in \mathbb{R}^2$ with
respective components $(p_1, \, p_2)$ and $(q_1, \, q_2)$.  Clearly
$\hat{\gamma}$ defines a cross-section of $J^1 \mathcal{B}$
first-order osculant to $\gamma$ at $p$ and so is a member of $j^1_q
[\gamma]$; indeed, it is the unique globally defined, linear
inhomogeneous map with this property.

A \emph{2-jet} is defined similarly, as an equivalence class of
sections under the relation of having the same first and second
partial-derivatives at a point.  More precisely, $\gamma, \eta \in
\Gamma [\mathcal{B}]$ \emph{osculate to second order} at $q \in
\mathcal{M}$ if $\gamma(q) = \eta(q)$ and
\begin{equation}
  \label{eq:coord-2jet}
  \begin{split}	
    \left. \frac{\partial (v^\alpha \circ \gamma)}{\partial x^i}
    \right|_q &= \left. \frac{\partial (v^\alpha \circ \eta)}{\partial
        x^i} \right|_q  \\
    \left. \frac{\partial^2 (v^\alpha \circ \gamma)}{\partial
        x^i \partial x^j} \right|_q &= \left. \frac{\partial^2
        (v^\alpha \circ \eta)}{\partial x^i \partial x^j} \right|_q
  \end{split}
\end{equation}
One then defines $J^2 \mathcal{B}$, \emph{et al}., in the analogous
ways.  There is a natural projection from $J^2 \mathcal{B}$ to $J^1
\mathcal{B}$, the \emph{truncation} $\theta^{2,1}$, characterized by
``dropping the second-order terms in the Taylor expansion''.  In
general, one has the natural truncation $\theta^{n,m} : J^n
\mathcal{B} \rightarrow J^m \mathcal{B}$ for all $0 < m < n$.

An important fact for the present goal of defining concomitants is
that the jet bundles of a geometric bundle are themselves naturally
geometric bundles.  Fix a geometric bundle $(\mathcal{B}, \,
\mathcal{M}, \pi, \, \iota)$ and a diffeomorphism $\phi$ on
$\mathcal{M}$.  Then $\iota[\phi]$ not only defines an action on
points of $\mathcal{B}$, but, as a diffeomorphism itself on
$\mathcal{B}$, it naturally defines an action on the cross-sections of
$\mathcal{B}$ and thus on the 1-jets.  The action can be characterized
by the appropriate coordinate transformations of the
formul{\ae}~\eqref{eq:coord-1jet} and \eqref{eq:coord-2jet}.  It is
easy to show that the mapping $\iota^1$ so specified from
$\mathfrak{D}_\mathcal{M}$ to $\mathfrak{D}^\sharp_{J^1 \mathcal{B}}$
is an injective homomorphism and thus itself an induction; therefore,
$(J^1 \mathcal{B}, \, \mathcal{M}, \sigma^1, \, \iota^1)$ is a
geometric fiber bundle.  One defines inductions for higher-order jet
bundles in the same way.

We can now generalize our definition of concomitants.  Let
$(\mathcal{B}_1, \, \mathcal{M}, \, \pi_1, \, \iota)$ and
$(\mathcal{B}_2, \, \mathcal{M}, \, \pi_2, \, \jmath)$ be two
geometric fiber bundles over the manifold $\mathcal{M}$.
\begin{defn}
  \label{defn:nth-concoms}
  An \emph{$n^{\text{th}}$-order concomitant} ($n$ a strictly positive
  integer) from $\mathcal{B}_1$ to $\mathcal{B}_2$ is a smooth mapping
  $\chi : J^n \mathcal{B}_1 \rightarrow \mathcal{B}_2$ such that
  \begin{enumerate}
      \item $( \forall u \in J^n \mathcal{B}_1 )( \forall \phi \in
    \mathfrak{A}_\mathcal{M} ) \; \jmath (\phi) (\chi(u)) = \chi (
    \iota^n (\phi) (u) )$
      \item there is no $(n-1)^{\text{th}}$ concomitant $\chi' :
    J^{n-1} \mathcal{B}_1 \rightarrow \mathcal{B}_2$ satisfying
    \[
    ( \forall u \in J^n \mathcal{B}_1 ) \; \chi (u) = \chi'
    (\theta^{n,n-1} (u))
    \]
  \end{enumerate}
\end{defn}
A zeroth-order concomitant (or just `concomitant' for short, when no
confusion will arise), is defined by \ref{def:concom}.  By another
convenient abuse of terminology, I will often refer to the range of
the concomitant mapping itself as `the concomitant' of the domain.  It
will be of physical interest in \S\ref{sec:conds} to consider the way
that concomitants interact with multiplication by a scalar field.
(Since we consider in this paper only concomitants of linear and
affine objects, multiplication of the object by a scalar field is
always defined.)  In particular, let us say that a concomitant is
\emph{homogeneous of weight $w$} if for any constant scalar field
$\xi$
\[
\chi (\iota_1 [\phi] (\xi u_1)) = \xi^w \iota_2 [\phi] (\chi (u_1))
\]

An important property of concomitants is that, in a limited sense,
they are transitive.
\begin{prop}
  \label{prop:con-composition}  
  If $\chi_1: J^n \mathcal{B}_1 \rightarrow \mathcal{B}_2$ is an
  $n^{\text{th}}$-order concomitant and $\chi_2: \mathcal{B}_2
  \rightarrow \mathcal{B}_3$ is a smooth mapping, where
  $\mathcal{B}_1$, $\mathcal{B}_2$ and $\mathcal{B}_3$ are geometric
  bundles over the same base space, then $\chi_2 \circ \chi_1$ is an
  $n^{\text{th}}$-order concomitant if and only if $\chi_2$ is a
  zeroth-order concomitant.
\end{prop}
This follows easily from the fact that inductions are injective
homomorphisms and concomitants respect the fibers.

\section{Concomitants of the Metric}
\label{sec:concomitants-metric}
\resetsec

As a specific example that will be of use in what follows, consider
the geometric fiber bundle $(\mathcal{B}_{\text{\small g}}, \,
\mathcal{M}, \, \pi_{\text{\small g}}, \, \iota_{\text{\small g}})$,
with $\mathcal{M}$ a 4-dimensional, Hausdorff, paracompact, connected,
smooth manifold (\emph{i}.\emph{e}., a candidate spacetime manifold),
the fibers of $\mathcal{B}_{\text{\small g}}$ diffeomorphic to the set
of Lorentz metrics at each point of $\mathcal{M}$, all of the same
signature $(+, \, -, \, -, \, -)$, and $\iota_{\text{\small g}}$ the
induction defined by the natural pull-back.  Since the set of Lorentz
metrics in the tangent plane over a point of a 4-dimensional manifold,
all of the same signature, is a 10-dimensional manifold,\footnote{In
  fact, it is diffeomorphic to a connected, convex, open subset---an
  open cone with vertex at the origin---in $\mathbb{R}^{10}$.} the
bundle space $\mathcal{B}_{\text{\small g}}$ is a 14-dimensional
manifold.  A cross-section of this bundle defines a Lorentz metric
field on the manifold.

The following proposition precisely captures the statement one
sometimes hears that there is no scalar or tensorial quantity one can
form depending only on the metric and its first-order partial
derivatives at a point of a manifold.
\begin{prop}
  \label{prop:no1metric}  
  There is no first-order concomitant from $\mathcal{B}_{\text{\small
      g}}$ to any tensor bundle over $\mathcal{M}$.
\end{prop}
To prove this, it suffices to remark that, given any spacetime
$(\mathcal{M}, \, g_{ab})$ and any two points $p,p' \in \mathcal{M}$,
there are coordinate neighborhoods $U$ of $p$ and $U'$ of $p'$ and a
diffeomorphism $\phi: \; \mathcal{M} \rightarrow \mathcal{M}$, such
that $\phi(p) = p'$, $\phi^\sharp (g'_{ab}) = g_{ab}$ at $p$, and
$\phi^\sharp(\partial'_a g_{bc}) = \partial_a g_{bc}$ at $p$, where
$\partial _a$ ($\partial'_a$) is the ordinary derivative operator
associated with the coordinate system on $U$ ($U'$), and $\phi^\sharp$
is the map naturally induced by the pull-back action of $\phi$.

This is not to say, however, that no information of interest is
contained in $J^1 \mathcal{B}_{\text{\small g}}$.  Indeed, two metrics
$g_{ab}$ and $h_{ab}$ are first-order osculant at a point if and only
if they have the same associated covariant derivative operator at that
point.  To see this, first note that, if they osculate to first order
at that point, then $\hat{\nabla}_a (g_{bc} - h_{bc}) = 0$ at that
point for all derivative operators.  Thus, for the derivative operator
$\nabla_a$ associated with, say, $g_{ab}$, $\nabla_a (g_{bc} - h_{bc})
= 0$, but $\nabla_a g_{bc} = 0$, so $\nabla_a h_{bc} = 0$ at that
point as well.  Similarly, if the two metrics are equal and share the
same associated derivative operator $\nabla_a$ at a point, then
$\hat{\nabla}_a (g_{bc} - h_{bc}) = 0$ at that point for all
derivative operators, since their difference will be identically
annihilated by $\nabla_a$, and $g_{ab} = h_{ab}$ at the point by
assumption.  Thus they are first-order osculant at that point and so
in the same 1-jet.  This proves that all and only geometrically
relevant information contained in the 1-jets of Lorentz metrics on
$\mathcal{M}$ is encoded in the fiber bundle over spacetime the values
of the fibers of which are ordered pairs consisting of a metric and
the metric's associated derivative operator at a spacetime point.

The second jet bundle over $\mathcal{B}_{\text{\small g}}$ has a
similarly interesting structure.  Clearly, if two metrics are in the
same 2-jet, then they have the same Riemann tensor at the point
associated with the 2-jet, since they have the same
partial-derivatives up to second order at the point.  Assume now that
two metrics are in the same 1-jet and have the same Riemann tensor at
the associated spacetime point.  If it follows that they are in the
same 2-jet, then essentially all and only geometrically relevant
information contained in the 2-jets of Lorentz metrics on
$\mathcal{M}$ is encoded in the fiber bundle over spacetime the points
of the fibers of which are ordered triplets consisting of a metric,
the metric's associated derivative operator and the metric's Riemann
tensor at a spacetime point.  To demonstrate this, it suffices to show
that if two Levi-Civita connections agree on their respective Riemann
tensors at a point, then the two associated derivative operators are
in the same 1-jet of the bundle whose base-space is $\mathcal{M}$ and
whose fibers consist of the affine spaces of derivative operators at
the points of $\mathcal{M}$ (because they will then agree on the
values of first-partial derivatives of their Christoffel symbols at
that point in any coordinate system as well as agreeing on the values
of the Christoffel symbols themselves, and thus will be in the 2-jet
of the same metric at that point).

Assume that, at a point $p$ of spacetime, $g_{ab} = \tilde{g}_{ab}$,
$\nabla_a = \tilde{\nabla}_a$ (the respective derivative operators),
and $R^a {}_{bcd} = \tilde{R}^a {}_{bcd}$ (the respective Riemann
tensors).  Let $C^a {}_{bc}$ be the symmetric difference-tensor
between $\nabla_a$ and $\tilde{\nabla}_a$, which is itself 0 at $p$ by
assumption.  Then by definition $\nabla_{[b} \nabla_{c]} \xi^a = R^a
{}_{bcn} \xi^n$ for any vector $\xi^a$, and so at $p$ by assumption
\begin{equation*}
  \begin{split}
    R^c {}_{abn} \xi^n &= \nabla_{[a} \tilde{\nabla}_{b]} \xi^c \\
    &= \nabla_a (\nabla_b \xi^c + C^c {}_{bn} \xi^n) -
    \tilde{\nabla}_b  \nabla_a\xi^c \\
    &= \nabla_a \nabla_b \xi^c + \nabla_a (C^c {}_{bn} \xi^n) -
    \nabla_b \nabla_a \xi^c - C^c {}_{bn} \nabla_a \xi^n + C^n {}_{ba}
    \nabla_n\xi^c
  \end{split}
\end{equation*}
but $\nabla_b \nabla_c \xi^a - \nabla_c \nabla_b \xi^a = 2 R^a
{}_{bcn} \xi^n$ and $C^a {}_{bc} = 0$, so expanding the only remaining
term gives
\[
\xi^n \nabla_a C^c {}_{bn} = 0
\]
for arbitrary $\xi^a$ and thus $\nabla_a C^b {}_{cd} = 0$ at $p$; by
the analogous computation, $\tilde{\nabla}_a C^b {}_{cd} = 0$ as well.
It follows immediately that $\nabla_a$ and $\tilde{\nabla}_a$ are in
the same 1-jet over $p$ of the affine bundle of derivative operators
over $\mathcal{M}$.  We have proved
\begin{theorem}
  \label{thm:1-2-jet-metric}
  $J^1 \mathcal{B}_{\text{\small g}}$ is naturally diffeomorphic to
  the geometric fiber bundle over $\mathcal{M}$ whose fibers consist
  of pairs $(g_{ab}, \, \nabla_a)$, where $g_{ab}$ is the value of a
  Lorentz metric field at a point of $\mathcal{M}$, and $\nabla_a$ is
  the value of the covariant derivative operator associated with
  $g_{ab}$ at that point, the induction being defined by the natural
  pull-back.  $J^2 \mathcal{B}_{\text{\small g}}$ is naturally
  diffeomorphic to the geometric fiber bundle over $\mathcal{M}$ whose
  fibers consist of triplets $(g_{ab}, \, \nabla_a, \, R_{abc} {}^d)$,
  where $g_{ab}$ is the value of a Lorentz metric field at a point of
  $\mathcal{M}$, and $\nabla_a$ and $R_{abc}{}^d$ are respectively the
  covariant derivative operator and the Riemann tensor associated with
  $g_{ab}$ at that point, the induction being defined by the natural
  pull-back.
\end{theorem}
It follows immediately that there is a first-order concomitant from
$\mathcal{B}_{\text{\small g}}$ to the geometric bundle
$(\mathcal{B}_\nabla, \, \mathcal{M}, \, \pi_\nabla,$ $\iota_\nabla)$
of derivative operators, \emph{viz}., the mapping that takes each
Lorentz metric to its associated derivative operator; likewise, there
is a second-order concomitant from $\mathcal{B}_{\text{\small g}}$ to
the geometric bundle $(\mathcal{B}_{\text{\small Riem}}, \,
\mathcal{M}, \, \pi_{\text{\small Riem}},$ $\iota_{\text{\small
    Riem}})$ of tensors with the same index structure and symmetries
as the Riemann tensor, \emph{viz}., the mapping that takes each
Lorentz metric to its associated Riemann tensor.  (This is the precise
sense in which the Riemann tensor associated with a given Lorentz
metric is ``a function of the metric and its partial derivatives up to
second order''.)  It is easy to see, moreover, that both concomitants
are homogeneous of degree 0.

It follows from theorem~\ref{thm:1-2-jet-metric} and
proposition~\ref{prop:con-composition} that a concomitant of the
metric of will be second order if and only if it is a zeroth-order
concomitant of the Riemann tensor, \emph{i}.\emph{e}., if and only if
it can be expressed as a linear combination of the Riemann tensor and
its contractions with the metric:
\begin{prop}
  \label{prop:0-concoms-riemann}
  A concomitant of the metric is second-order if and only if it can be
  expressed as a sum of terms consisting of constants multiplied by
  the Riemann tensor, the Weyl tensor, the Ricci tensor, the Gaussian
  scalar curvature, and contractions and products of these with the
  metric itself.
\end{prop}

\section{Conditions on a Possible Gravitational Stress-Energy Tensor}
\label{sec:conds}
\resetsec

We are almost in a position to state and prove the main result of the
paper, the nonexistence of a gravitational stress-energy tensor.  In
order to formulate and prove a result having that proposition as its
natural interpretation, one must first lay down some natural
conditions on the proposed object, to show that no such object exists
satisfying the conditions.  In general relativity, the invariant
representation of energetic quantities is always in the form of a
stress-energy tensor, \emph{viz}., a two-index, symmetric,
divergence-free tensor.  Not just any such tensor will do, however,
for that gives only the baldest of formal characterizations of it.
From a physical point of view, at a minimum the object must have the
physical dimension of stress-energy for it to count as a stress-energy
tensor.  That it have the dimension of stress-energy is what allows
one to add two of them together in a physically meaningful way to
derive the physical sum of total stress-energy from two different
sources.  In classical mechanics, for instance, both velocity and
spatial position have the form of a three-dimensional vector, and so
their formal sum is well defined, but it makes no physical sense to
add a velocity to a position because the one has dimension
\texttt{length/time} and the other the dimension \texttt{length}.  (I
will give a precise characterization of ``physical dimension'' below.)

An essential, defining characteristic of energy in classical physics
is its obeying some formulation of the First Law of Thermodynamics.
The formulation of the First Law I rely on is somewhat unorthodox:
that all forms of stress-energy are in principle ultimately
fungible---any form of energy can in principle be transformed into any
other form\footnote{\citeN[ch.~\textsc{v}, \S97]{maxwell-matt-mot}
  makes this point especially clearly.}---not necessarily that there
is some absolute measure of the total energy contained in a system or
set of systems that is constant over time.  In more precise terms,
this means that all forms of energy must be represented by
mathematical structures that allow one to define appropriate
operations of addition and subtraction among them, which the canonical
form of the stress-energy does allow for.\footnote{Note that this is a
  requirement even if one takes a more traditional view of the First
  Law as making a statement about conservation of a magnitude
  measuring a physical quantity.}  I prefer this formulation of the
First Law in general relativity because there will not be in a general
cosmological context any well-defined global energetic quantity that
one can try to formulate a conservation principle for.  In so far as
one wants to hold on to some principle like the classical First Law in
a relativistic context, therefore, I see no other way of doing it
besides formulating it in terms of fungibility.  (If one likes, one
can take the fungibility condition as a necessary criterion for any
more traditional conservation law.)  This idea is what the demand that
\emph{all} stress-energy tensors, no matter the source, have the same
physical dimension is intended to capture.\footnote{For what it's
  worth, this conception has strong historical warrant---Einstein
  (implicitly) used a very similar formulation in one of his first
  papers laying out and justifying the general theory
  \cite[p.~149]{einstein16/52}:
  \begin{quote}
    It must be admitted that this introduction of the energy-tensor of
    matter is not justified by the relativity postulate alone.  For
    this reason we have here deduced it from the requirement that the
    energy of the gravitational field shall act gravitatively in the
    same way as any other kind of energy.
  \end{quote}
  \citeN{moller-energy-mom-gr} also stresses the fact that the
  formulation of integral conservation laws in general relativity
  based on pseudo-tensorial quantities depends crucially on the
  assumption that gravitational energy, such as it is, shares as many
  properties as possible with the energy of ponderable
  (\emph{i}.\emph{e}., non-gravitational) matter.}

To sum up, the stress-energy tensor encodes in general relativity all
there is to know of ponderable (\emph{i}.\emph{e}., non-gravitational)
energetic phenomena at a spacetime point:
\begin{enumerate}  
    \item it has 10 components representing with respect to a fixed
  pseudo-orthonormal frame, say, the 6 components of the classical
  stress-tensor, the 3 components of linear momentum and the scalar
  energy density of the ponderable field at that point
    \item that it has two covariant indices represents the fact that
  it defines a linear mapping from timelike vectors at the point
  (``worldline of an observer'') to covectors at that point
  (``4-momentum covector of the field as measured by that observer''),
  and so defines a bi-linear mapping from pairs of timelike vectors to
  a scalar density at that point (``scalar energy density of the field
  as measured by that observer''), because energetic phenomena,
  crudely speaking, are marked by the fact that they are quadratic in
  velocity and momental phenomena linear in velocity
    \item that it is symmetric represents, ``in the limit of the
  infinitesimal'', the classical principle of the conservation of
  angular momentum; it also encodes part of the relativistic
  equivalence of momentum-density flux and scalar energy density
    \item that it is covariantly divergence-free represents the fact
  that, ``in the limit of the infinitesimal'', the classical
  principles of energy and linear momentum conservation are obeyed; it
  also encodes part of the relativistic equivalence of
  momentum-density flux and scalar energy density
    \item the localization of ponderable stress-energy and its
  invariance as a physical quantity are embodied in the fact that the
  object representing it is a \emph{tensor}, a multi-linear map acting
  only on the tangent plane of the point it is associated with
    \item finally, the thermodynamic fungibility of energetic
  phenomena is represented by the fact that the set of stress-energy
  tensors forms a vector space---the sum and difference of any two is
  itself a possible stress-energy tensor---all elements of which have
  the same physical dimension
\end{enumerate}
Consequently, the appropriate mathematical representation of localized
gravitational stress-energy, if there is such a thing, is a two
covariant-index, symmetric, covariantly divergence-free tensor having
the physical dimension of stress-energy.  (That we demand it be
covariantly divergence-free is a delicate matter requiring special
treatment, which I give at the end of this section.)

Now, in order to make precise the idea of having the physical
dimension of stress-energy, recall that in general relativity all the
fundamental units one uses to define stress-energy, namely time,
length and mass, can themselves be defined using only the unit of
time; these are so-called geometrized units.  For time, this is
trivially true: stipulate, say, that a time-unit is the time it takes
a certain kind of atom to vibrate a certain number of times under
certain conditions.  A unit of length is then defined as that in which
light travels \emph{in vacuo} in one time-unit.  A unit of mass is
defined as that of which two, placed one length-unit apart, will
induce in each other by dint of their mutual gravitation alone an
acceleration towards each other of one length-unit per time-unit per
time-unit.\footnote{This definition may appear circular, in that it
  would seem to require a unit of mass in the first place before one
  could say that bodies were of the \emph{same} mass.  I think the
  circularity can be mitigated by using two bodies for which there are
  strong prior grounds for positing that they are of equal mass,
  \emph{e}.\emph{g}., two fundamental particles of the same type.  It
  also suffers from a fundamental lack of rigor that the definition of
  length does not suffer from.  In order to make the definition
  rigorous, one would have to show, \emph{e}.\emph{g}., that there
  exists a solution of the Einstein field-equation (approximately)
  representing two particles in otherwise empty space (as defined by
  the form of $T_{ab}$)---\emph{viz}., two timelike geodesics---such
  that, if on a spacelike hypersurface at which they both intersect 1
  unit of length apart (as defined on the hypersurface with respect to
  either) they accelerate towards each other (as defined by relative
  acceleration of the geodesics) one unit length per unit time
  squared, then the product of the masses of the particles is 1.  I
  will just assume, for the purposes of this paper, that such
  solutions exist.}  These definitions of the units of mass and length
guarantee that they scale in precisely the same manner as the
time-unit when new units of time are chosen by multiplying the
time-unit by some fixed real number $\lambda^{-\frac{1}{2}}$.  (The
reason for the inverse square-root will become clear in a moment).
Thus, a duration of $t$ time-units would become
$t\lambda^{-\frac{1}{2}}$ of the new units; an interval of $d$ units
of length would likewise become $d\lambda^{-\frac{1}{2}}$ in the new
units, and $m$ units of mass would become $m\lambda^{-\frac{1}{2}}$ of
the new units.  This justifies treating all three of these units as
``the same'', and so expressing acceleration, say, in inverse
time-units.  To multiply the length of all timelike vectors
representing an interval of time by $\lambda^{-\frac{1}{2}}$, however,
is equivalent to multiplying the metric by $\lambda$ (and so the
inverse metric by $\lambda^{-1}$), and indeed such a multiplication is
the standard way one represents a change of units in general
relativity.  This makes physical sense as the way to capture the idea
of physical dimension: all physical units, the ones composing the
dimension of any physical quantity, are geometrized in general
relativity, in the most natural formulation, and so depend only on the
scale of the metric itself.

Now, the proper dimension of a stress-energy tensor can be determined
by the demand that the Einstein field-equation, $G_{ab} = \gamma
T_{ab}$, where $\gamma$ is Newton's gravitational constant, remain
satisfied when one rescales the metric by a constant factor.  $\gamma$
has dimension $\frac{\mbox{(length)}^3} {\mbox{(mass)(time)}^2}$, and
so in geometrized units does not change under a constant rescaling of
the metric.  Thus $T_{ab}$ ought to transform exactly as $G_{ab}$
under a constant rescaling of the metric.  A simple calculation shows
that $G_{ab}$ $(= R_{ab} - \half R g_{ab})$ remains unchanged under
such a rescaling.  Thus, a necessary condition for a tensor to
represent stress-energy is that it remain unchanged under a constant
rescaling of the metric.  It follows that the concomitant at issue
must be homogeneous of weight 0 in the metric, whatever order it may
be.

We must still determine the order of the required concomitant, for it
is not \emph{a priori} obvious.  In fact, the way a homogeneous
concomitant of the metric transforms when the metric is multiplied by
a constant factor suffices to fix the differential order of that
concomitant.\footnote{I thank Robert Geroch for pointing this out to
  me.}  This can be seen as follows, as exemplified by the case of a
two covariant-index, homogeneous concomitant $S_{ab}$ of the metric.
A simple calculation based on definition~\ref{defn:nth-concoms} and on
the fact that the concomitant must be homogeneous shows that the value
at a point $p \in \mathcal{M}$ of an $n^{\text{th}}$-order concomitant
$S_{ab}$ can be written in the general form
\begin{equation}
  \label{eq:Sab-form}
  S_{ab} = \sum_\alpha k_\alpha \space g^{qx} \ldots g^{xr} \left(
    \widetilde{\nabla}_x^{(n_1)} g_{qx} \right) \ldots \left(
    \widetilde{\nabla}_x^{(n_i)} g_{xr} \right)
\end{equation}
where: $\widetilde{\nabla}_a$ is any derivative operator at $p$
\emph{other} than the one naturally associated with $g_{ab}$; `$x$' is
a dummy abstract index; `$\widetilde{\nabla}_x^{(n_i)}$' stands for
$n_i$ iterations of that derivative operator (obviously each with a
different abstract index); $\alpha$ takes its values in the set of all
permutations of all sets of positive integers $\{ n_1, \ldots, n_i \}$
that sum to $n$, so $i$ can range in value from 1 to $n$; the
exponents of the derivative operators in each summand themselves take
their values from $\alpha$, \emph{i}.\emph{e}., they are such that
$n_1 + \cdots + n_i = n$; there is exactly one summand for which $n_1
= n$ (which makes it an $n^{\text{th}}$-order concomitant); for each
$\alpha$, $k_\alpha$ is a constant; and there are just enough of the
inverse metrics in each summand to contract all the covariant indices
but $a$ and $b$.

Now, a combinatorial calculation shows
\begin{prop}
  \label{prop:nth-concom-factor}
  If, for $n \geq 2$, $S_{ab}$ is an $n^{\text{th}}$-order homogeneous
  concomitant of $g_{ab}$, then to rescale the metric by the constant
  real number $\lambda$ multiplies $S_{ab}$ by $\lambda^{n - 2}$.
\end{prop}
In other words, the only such homogeneous $n^{\text{th}}$-order
concomitants must be of weight $\lambda - 2$.\footnote{Note that the
  exponent $(n - 2)$ in this result depends crucially on the fact that
  $S_{ab}$ has only two indices, both covariant.  One can generalize
  the result for tensor concomitants of the metric of any index
  structure.  A slight variation of the argument, moreover, shows that
  there does not in general exist a homogeneous concomitant of a given
  order from a tensor of a given index structure to one of another
  structure---one may not be able to get the number and type of the
  indices right by contraction and tensor multiplication alone.}  So
if one knew that $S_{ab}$ were multiplied by, say, $\lambda^4$ when
the metric was rescaled by $\lambda$, one would know that it had to be
a sixth-order concomitant.  In particular, $S_{ab}$ does not rescale
when $g_{ab} \rightarrow \lambda g_{ab}$ only if it is a second-order
homogeneous concomitant of $g_{ab}$, \emph{i}.\emph{e}., (by
theorem~\ref{thm:1-2-jet-metric} and
proposition~\ref{prop:0-concoms-riemann}) a zeroth-order concomitant
of the Riemann tensor.  There follows from
proposition~\ref{prop:con-composition}
\begin{lemma}
  \label{lem:riem-0th-concom-0-homog}
  A 2-covariant index concomitant of the Riemann tensor is homogeneous
  of weight zero if and only if it is a zeroth-order concomitant.
\end{lemma}
Thus, such a tensor has the physical dimension of stress-energy if and
only if it is a zeroth-order concomitant of the Riemann
tensor.

We now address the issue whether it is appropriate to demand of a
potential gravitational stress-energy tensor that it be covariantly
divergence-free.  In general, I think it is not, even though that is
one of the defining characteristics of the stress-energy tensor of
ponderable matter in the ordinary formulation of general
relativity.\footnote{I thank David Malament for helping me get
  straight on this point.  The following argument is in part
  paraphrastically based on a question he posed to me.}  To see this,
let $T_{ab}$ represent the aggregate stress-energy of all ponderable
matter fields.  Let $S_{ab}$ be the gravitational stress-energy
tensor, which we assume for the sake of argument to exist.  Now, we
ask: can the ``gravitational field'' interact with ponderable matter
fields in such a way that stress-energy is exchanged?  If it could,
then, presumably, there could be interaction states characterized (in
part) jointly by these conditions:
\begin{enumerate}
    \item $\nabla^n (T_{na} + S_{na}) = 0$
    \item $\nabla^n T_{na} \ne 0$
    \item $\nabla^n S_{na} \ne 0$
\end{enumerate}
The most one can say, therefore, without wading into some very deep
and speculative waters about the way that a gravitational
stress-energy tensor (if there were such a thing) might enter into the
righthand side of the Einstein field-equation, is that we expect such
a thing would have vanishing covariant divergence when the aggregate
stress-energy tensor of ponderable matter vanishes,
\emph{i}.\emph{e}., that gravitational stress-energy on its own, when
not interacting with ponderable matter, be conserved.  This weaker
statement will suffice for our purposes, so we can safely avoid those
deep waters.

Finally, it seems reasonable to require one more condition: were there
a gravitational stress-energy tensor, it should not be zero in any
spacetime with non-trivial curvature, for one can always envision the
construction of a device to extract energy in the presence of
curvature by the use of tidal forces and geodesic
deviation.\footnote{See, \emph{e}.\emph{g}., \citeN{bondi62}.}

To sum up, we have the following necessary condition:
\begin{condition}
  \label{cond:candidates-sab}
  The only viable candidates for a gravitational stress-energy tensor
  are two covariant-index, symmetric, second-order, zero-weight
  homogeneous concomitants of the metric that are not zero when the
  Riemann tensor is not zero and that have vanishing covariant
  divergence when the stress-energy tensor of ponderable matter
  vanishes.
\end{condition}
This discussion, by the way, has obviated the criticism of the claim
that gravitational stress-energy ought to depend on the curvature,
\emph{viz}., that this would make gravitational stress-energy depend
on second-order partial derivatives of the field potential whereas all
other known forms of stress-energy depend only on terms quadratic in
the first partial derivatives of the field potential.  It is exactly
second-order, homogeneous concomitants of the metric that possess
terms quadratic in the first partials.  The rule is that the order of a
homogeneous concomitant is the sum of the exponents of the derivative
operators when the concomitant is represented in the form of
equation~(\ref{eq:Sab-form}).

\section{There Is No Gravitational Stress-Energy Tensor}
\label{sec:nonexist}
\resetsec

It follows from lemma~\ref{lem:riem-0th-concom-0-homog}, in
conjunction with condition~\ref{cond:candidates-sab}, that any
candidate gravitational stress-energy tensor must be a zeroth-order
concomitant of $\mathcal{B}_{\mbox{\small Riem}}$, the geometric
bundle of Riemann tensors over spacetime.  (One can take this as a
precise statement of the fact that any gravitational stress-energy
tensor ought to ``depend on the curvature'', as I argued in
\S\ref{sec:princ_equiv}.)  From
proposition~\ref{prop:0-concoms-riemann} and Lovelock's theorem stated
in footnote~\ref{fn:schouten-lovelock}, it follows that the only
possibilities then are constant multiples of the Einstein tensor, but
that tensor can still be zero even when the Riemann tensor is not
(when, \emph{e}.\emph{g}., there is only Weyl curvature).  This proves
the main result.
\begin{theorem}
  \label{thm:non-exist}
  There are no two covariant-index, symmetric, divergence-free,
  second-order, homogeneous concomitants of the metric that are not
  zero when the Riemann tensor is not zero.
\end{theorem}
The theorem does bear the required natural interpretation, for the
Einstein tensor is not an appropriate candidate for the representation
of gravitational stress-energy: the Einstein tensor will be zero in a
spacetime having a vanishing Ricci tensor but a non-trivial Weyl
tensor; such spacetimes, however, can manifest phenomena,
\emph{e}.\emph{g}., pure gravitational radiation in the absence of
ponderable matter, that one naturally wants to say possess
gravitational energy in some (necessarily non-localized) form or
other.\footnote{As an historical aside, it is interesting to note that
  early in the debate on gravitational energy in general relativity
  \citeN{lorentz-zwaartekracht-3} and \citeN{levi-civita-grav-tensor}
  proposed that the Einstein tensor be thought of as the gravitational
  stress-energy tensor.  Einstein criticized the proposal on the
  grounds that this would result in attributing zero total energy to
  any closed system.}

\end{document}